\documentclass[preprint,aps,draft]{revtex4}
\usepackage{graphicx}
\usepackage{dcolumn}
\usepackage{bm}
\usepackage{color}

\newcommand{\be}{\begin{equation}}
\newcommand{\ee}{\end{equation}}
\newcommand{\bea}{\vspace{0.25cm}\begin{eqnarray}}
\newcommand{\eea}{\end{eqnarray}}

\def\PRL{{Phys. Rev. Lett.} }

\begin{document}

\centerline{ \large \bf Conditional Unitary Transformation on
biphotons} \vskip 1.2cm \centerline{  G. Brida$^1$, M.V.
Chekhova$^2$, M. Genovese$^1$, M. Gramegna$^1$, L.A.
Krivitsky$^2$, S.P.Kulik$^2$} \vskip 0.5cm \centerline{ \it $^1$
Istituto Elettrotecnico Nazionale Galileo Ferraris,} \centerline{
\it Strada delle Cacce 91, 10135 Torino, Italy } \centerline{ \it
$^2$ Physics Department, M.V. Lomonosov Moscow State University,
119992 Moscow, Russia.}

\centerline{ \bf Abstract} \vskip 0.2cm

A conditional unitary transformation ($90^o$ polarization
rotation) is performed at single-photon level. The transformation
is realized by rotating polarization for one of the photons  of a
polarization-entangled biphoton state (signal photon) by means of
a Pockel cell triggered by the detection of the other (idler)
photon after polarization selection. As a result, the state of the
signal photon is losslessly changed from being completely
unpolarized to being partially polarized, so that the final
polarization degree is given by the idler detector quantum
efficiency. This experiment can be used for developing a novel
method of absolute quantum efficiency calibration.


\vskip 1cm

The possibility of performing a unitary transformation on a qubit
conditionally on the result of the measurement on another one is a
fundamental tool for quantum information \cite{nc}. For example,
in teleportation protocols \cite{teleth}, teleportation is
performed by Bob doing an opportune unitary transformation on his
sub-part of an entangled system after having received  classical
information on the result of the Bell measurement performed by
Alice on her sub-part of the entangled system and the unknown
state to be teleported. Similarly, in entanglement swapping
protocols a joint measurement performed on two members of two
distinct entangled pairs allows, by means of a conditional unitary
transformation, to create a desired entangled state between the
two surviving members.
 In most of experiments realized up to now with
 single photons both for teleportation \cite{tele} and swapping
 \cite{swex}, this part of the
 protocol was not accomplished, and only correlation measurements with a fixed polarization selection
 were obtained. The only exception is~\cite{deMartini}, where a Pockel cell was used
 for conditional transformations in an active teleportation protocol. Conditional transformations were
 also performed in experiments where one of the photons of an entangled pair
triggered polarization rotation on the other one, to increase the
efficiency of the parity
 check~\cite{pitt1} and to produce signal photons on 'pseudodemand', i.e., in a given time interval
 after the trigger detection event~\cite{pitt2}.
Single-photon light was observed through the anti-bunching effect
in~\cite{Rarity} as well, following a proposal of \cite{Klyshko}
where it was suggested to prepare pure single-photon state by
gating (using a shutter) the signal photon of an entangled pair as
soon as the idler photon is detected.
 Finally, for the sake of completeness, it can still be acknowledged that, somewhat later,
 it was proposed to produce sub-Poissonian light
 by means of conditional transformations on twin beams \cite{yamamoto}.
 Conditional preparation of sub-Poissonian light from twin beams was
 carried out in experiments using active control (feedforward)~\cite{fabre1} and passive measurement
 technique~\cite{fabre2}.

The idea of the present work is to perform conditional
 unitary transformations on the polarization state of a photon
 belonging to a polarization-entangled pair. This experiment
 is the first demonstration of 'purifying' the polarization state of a photon that is
 initially prepared in a
 mixed polarization state
 (initial degree of polarization is equal to zero).
 As a result of transformations, the degree of polarization increases; its final value is given
 by the quantum efficiency of the 'trigger' detector. Due to this
 fact, one can notice another important advantage of our experiment: it can be used for the absolute calibration of
 photodetectors (absolute measurement of quantum efficiency).

In our experiment, biphotons are generated via spontaneous
parametric down conversion (SPDC) by pumping a  BBO crystal (5x5x5
mm) with an argon laser radiating at 351 nm. The crystal is cut
for type-II non-collinear frequency-degenerate phase matching
providing generation of polarization-wavevector entangled
state~\cite{zei}. In this regime, signal and idler photons are
emitted in different directions and each of them is not polarized;
at the same time, their polarization states are always orthogonal.
The state at the output of the crystal has the form

\begin{equation}
| \psi \rangle = \frac{ \vert H \rangle \vert V \rangle +
\rm{e}^{i\phi}\vert V \rangle \vert H \rangle }{\sqrt{2}},
\label{psi}
\end{equation}
where $H,V$ denote horizontal and vertical polarization,
respectively, and the first and the second positions correspond to
the two spatial modes (two directions). The phase $\phi$ can be
varied by small tilts of the crystal and by additionally
introduced optic elements.

The state (1) is a pure entangled state. Strictly speaking, in
order to obtain this state, it is necessary that the group delay
between vertically and horizontally polarized photons is
compensated after the crystal~\cite{comp}. As a compensator, one
can use a BBO crystal of length twice less than for the one where
photon pairs are generated, or any other crystal of equivalent
length. Without this compensation, the phase $\phi$ in Eq. (1) has
classical random fluctuations or, in other words, the state
produced at the output is a mixed entangled state, described by
the density matrix \be \rho = {\frac{ \vert HV\rangle \langle HV
\vert + \vert VH \rangle \langle VH \vert }{2}}. \label{rho} \ee

In our experiment, no compensator was used, and hence, a mixed
entangled state (2) was generated. Note that each of the photons
forming the entangled pair is in a mixed polarization state, no
matter whether the two-photon state is pure (Eq.(1)) or mixed
(Eq.(2)). The effect of 'purification' observed in the present
work for the polarization state of the signal photon does not
require the purity of the two-photon state. Some additional
features should be observed in the case where the two-photon state
is pure, but this is not the subject of the present paper and will
be considered elsewhere.

After the BBO crystal, the pump was absorbed by a beam trap. One
of the correlated photons (the idler photon) was sent to a
single-photon counter (Perkin-Elmer SPCM-AQ) preceded by a
red-glass cutoff filter and a pinhole (realizing a rough spectral
and spatial selection) and then to a polarizer cube selecting
vertically polarized photons. The output of this detector was sent
to a counter and, as a trigger, to a pulse generator used to
control a KDP Pockel cell placed on the path of the conjugated
(signal) photon. The Pockel cell was operating as a halfwave plate
oriented at 45$^{\circ}$ to the vertical axis (thus providing
polarization rotation by 90$^{\circ}$) if supplied by the 5.2 kV
peak voltage from the pulse generator. The high-voltage pulse had
a sharp front, not longer than 2 ns, a flat part of duration 100
ns, and a 'tail' 2$\mu \textrm{s}$ long. To ensure that the flat
part of the high-voltage pulse arrived at the Pockel cell
simultaneously with the signal photon, signal photons were delayed
by means of 50 m of polarization-maintaining (PM) fibre. For
additional adjustment, a variable delay between the idler detector
output pulse and the high-voltage pulse could be introduced
electronically. This way, each detection of an idler photon with
vertical polarisation led to a unitary operation (polarisation
rotation by 90$^{\circ}$) on the corresponding signal photon.
Then, after an interference filter with 4 nm FWHM centered at 702
nm and a Glan-Thompson polarizer, the signal photon was detected
by a single-photon counter (Perkin-Elmer SPCM-AQ).

When the Pockel cell is not active, the reduced polarization
density matrix of the signal photon corresponds to a completely
unpolarized case. In fact, the observed dependence of the signal
beam intensity on the setting of the Glan-Thompson polarizer was
flat (see Fig.2).

 On the other hand, when the Pockel
cell is active, the system realizes a unitary transformation
(rotation by $90^o$) on the polarization of the signal photon
conditioned to the registration of a vertically polarized idler
photon. The polarization density matrix of the signal photon
corresponds therefore to a perfectly vertically polarized state
(if  the idler detector has perfect quantum efficiency and the
Pockel cell operates in 100\% cases): the initially perfectly
mixed state of the signal photon $1/2 (\vert V \rangle \langle V
\vert + \vert H \rangle \langle H \vert)$ is purified to the state
$\vert V \rangle \langle V \vert$. In a real situation, the
quantum efficiency $\eta$ of the first detector is smaller than
unity, which means that only a fraction of the incident photons
will be observed and produce an effect on the Pockel cell. Thus,
the final density matrix for the second photon is \be \rho =1/2 (
(1+\eta) \vert V \rangle \langle V \vert + (1- \eta) \vert H
\rangle \langle H \vert) \ee

It is important that the transformation performed on the signal
photon, in principle, is lossless, since the only sources of
losses are reflection and absorption in the optical elements,
which can be made as small as possible.

This effect can be also described in terms of the Stokes
parameters, which are initially $S_0=N$, $S_1=S_2=S_3=0$, where
$N$ is the photon number, but become, as a result of
transformation, $S_0=N$, $S_1= \eta N$, $S_2=S_3=0$. From the
Stokes parameters, one can find the polarization degree, which is
standardly defined as \be
P=\sqrt{\frac{S_1^2+S_2^2+S_3^2}{S_0^2}}.\ee We see that without
the transformation, $P=0$ and in the presence of the
transformation, $P=\eta$. Since the degree of polarization depends
on $\eta$ this scheme could also be used for an absolute
calibration of detectors (a detailed description of this technique
will appear in Ref. \cite{nos}).

 The results of our measurements
 are shown in Figs.2, 3. As expected, in the
absence of the Pockel cell action, the measured flux on the signal
detector does not depend on the polarization selection. On the
other hand, when the cell is active the measured signal has a $(1+
\eta cos (2 \theta))$ behaviour as a function of the polarizer
setting $\theta$, with a $30 \%$ visibility (Fig.2). Subtracting
the noise background of the signal detector and correcting for
Pockel cell system inefficiency, we discover that the visibility
becomes $0.441 \pm 0.045$, which is close to the result
$\eta=0.476 \pm 0.002$ obtained by means of a standard biphoton
calibration scheme \cite{bp1,bp2}. The dependence of coincidence
counting rate on the position of the polarizer in the signal
channel is shown in Fig.3. In contrast to the single-photon
counting rate, this dependence should not be sensitive to the
idler detector quantum efficiency. The effect of the Pockel cell
manifests itself in the $90^{\circ}$ phase shift of the
dependence. A small decrease in the visibility (from $95\%$ to
$90\%$) can be explained only by 'failures' of the Pockel cell
caused by the effect of its dead time (2 $\mu$s).

To demonstrate that the effect occurs only in cases where the
polarization transformation on the signal photon is triggered by
its conjugated idler photon, we studied the dependence of the
effect on the trigger pulse delay $T$ introduced electronically.
To obtain this dependence, the electronic delay was scanned from 0
to 200 ns while the polarizer in the signal channel was fixed in
the positions $\theta=0^{\circ}$ and $\theta=90^{\circ}$ (Fig.4).
The dependence shows that at zero electronic delay, the effect is
observed but at delays larger than $100$ ns, the effect
disappears. To explain this dependence, we have to mention that
the electronic part of the setup has a constant uncontrollable
delay $T_0$ introduced by the pulse generator. At zero additional
delay $T$, the effect is present because the delay $T_f$
introduced by the fibre ($248$ ns, according to our measurement)
is small enough compared to the delay $T_0$ plus the duration of
the high-voltage pulse. Disappearance of the effect at $T>100$ ns
occurs because in this case $T+T_0$ starts to exceed the delay
$T_f$ introduced by the fibre. By taking a longer fibre we could
observe no effect at small $T$ and at larger $T$, the initial part
of the dependence would show the 'tail' of the high-voltage pulse;
however, this tail is more than one microsecond long and the PM
fibre does not maintain polarization well enough at lengths larger
than several hundreds of meters.

In conclusion, we have performed a conditional unitary
transformation at single-photon level. The state of the signal
photon, which was initially mixed, became partly purified due to
this transformation. In terms of polarization properties, light
became partly polarized (polarization degree became equal to the
quantum efficiency of the trigger detector) while initially, it
was completely non-polarized (polarization degree was zero). Note
that the transformation is lossless with respect to the signal
photon. It is worth mentioning that variation of the polarization
degree, which we observe in our experiment, is impossible by means
of only linear lossless optical methods~\cite{pd}. However, in the
present experiment, the transformation performed over the signal
photon is essentially nonlinear since it is triggered by the
detection of the idler photon entangled to this photon.

\vskip 0.5cm

{\bf Acknowledgments}

We acknowledge support of INTAS, grant \#01-2122. One of us (L.K.)
acknowledges support from INTAS-YS fellowship grant (Num.
03-55-1971). The Turin group acknowledges the support of: MIUR
(FIRB RBAU01L5AZ-002; Cofinanziamento 2001) and Regione Piemonte.
The Moscow group acknowledges support of the Russian Foundation
for Basic Research, grant\#02-02-16664, and the Russian program of
scientific school support (\#166.2003.02).
 \vskip 3cm


\newpage
\centerline{ \bf Figure Captions}

Fig.1. The experimental set-up. An argon laser generating at 351
nm pumps a type-II BBO crystal cut for frequency-degenerate
non-collinear polarization-entangled phase matching. One of the
correlated photons, after a spatial selection by means of an
aperture A, a spectral selection by means of a red-glass cutoff
filter RG, and a polarization selection by means of a polarizing
cube PBS, is addressed to the photon counter D1, which drives,
through a fast high-voltage switch S, a Pockel cell PC placed in
the optical path of the other photon. The delay between a
photocount of D1 and the corresponding high-voltage pulse on the
Pockel cell can be varied electronically. The second photon of an
entangled pair is retarded, before the Pockel cell, by means of a
50 m PM fibre F. This realizes the conditioned unitary operation.
The second photon is registered by photon counter D2 preceded by a
Glan prism G and an interference filter IF. The measurement
includes registration of the D2 single-photon counting rate and
the rate of coincidences between D1 and D2.

Fig.2.  Counting rate of detector D2 as a function of the angle of
the polarizer preceding it.  When the Pockel cell is not activated
no dependence on the polarizer angle appears (squares). When a
$90^{\circ}$ rotation of polarization is realized by the Pockel
cell conditioned to the measurement of a vertically polarized
photon in the conjugated branch, the data (triangles) show a clear
dependence on the polarizer setting corresponding to a vertically
polarized state, with a  $30 \%$ visibility~\cite{subtraction}.

Fig.3. Coincidences between D1 and D2 as a function of the
polarizer orientation. The data when the Pockel cell is activated
show a $90 \%$ visibility. The reduction of visibility can be
attributed to a non-perfect efficiency of the Pockel cell
apparatus.

Fig.4.  Dependence of the D2 counting rate at $\theta=0^o$
(Horizontally polarized photons, triangles) and $\theta=90^o$
(Vertically polarized photons, squares) on the delay $T$
introduced electronically between the trigger pulses from D1
detector and the corresponding high-voltage pulses driving the
Pockel's cell. The effect of counting rate decreasing for
horizontally polarized photons is observed for  $T < 100 ns$ and
not observed for $T>100 ns$.

  \vskip 0.5cm


\begin{thebibliography}{99}
\bibitem{subtraction} After subtracting the
noise counts we obtain the visibility of 44\%.
\end{thebibliography}
\end{document}